\documentclass[usenatbib]{mn2e}

\usepackage{amsmath}
\usepackage{amssymb}
\usepackage{graphicx}

\newcommand{\rst}{\rho^*}

\newcommand{\pa}[1]{\partial_#1}
\newcommand{\pau}[1]{\partial^#1}

\def\agt{\mathrel{\raise.3ex\hbox{$>$}\mkern-14mu\lower0.6ex\hbox{$\sim$}}}
\def\alt{\mathrel{\raise.3ex\hbox{$<$}\mkern-14mu\lower0.6ex\hbox{$\sim$}}}

\author[R. Oechslin et al.]{R. Oechslin$^1$\thanks{present address: Max-Planck Institut f\"ur Astrophysik,
Karl Schwarzschild-Str. 1, 85741 Garching, Germany}, K. Ury\=u$^2$,
G. Poghosyan$^1$, F. K. Thielemann$^1$\\
$^1$Departement f\"ur Physik und Astromonie der
Universit\"at Basel, Klingelbergstr. 82, CH-4056 Basel, Switzerland\\
$^2$SISSA, Via Beirut 2/4, 34014 Trieste, Italy}

\title[Binary neutron star mergers]{
The Influence of Quark Matter at High Densities on 
Binary Neutron Star Mergers}

\begin{document}
\maketitle
\begin{abstract}
We consider the influence of potential quark matter existing at high
densities in neutron star interiors on gravitational waves (GW) 
emitted in a binary neutron star merger event.  
Two types of equations of state (EoS) at zero temperatures are used, 
one describing pure nuclear matter, the other nuclear
matter with a phase transition to quark matter at very high
densities. 
Binary equilibrium sequences close to the
innermost stable circular orbit (ISCO) are calculated to determine the
GW frequencies just before merger. 
It is found that EoS effects begin to play a role for gravitational masses
larger than $M_\infty\simeq1.5M_\odot$. The difference in the
gravitational wave frequency at the ISCO grows to up to $\simeq 10\%$
for the maximal allowed mass given by the EoSs used.
Then, we perform 3D hydrodynamic simulations 
for each EoS varying the initial mass and determine the characteristic 
GW frequencies of the merger remnants.
The implications of quark matter show up mainly in a different collapse
behaviour of the merger remnant. If the collapse does not take place
immediately after merger, we find a phase difference between two 
EoS's in the post-merger GW signal.
We also compare the GW frequencies emitted by the merger remnant to 
values from simulations using a polytropic EoS and find an imprint of the
non-constant adiabatic index of our EoSs.
All calculations are based on the conformally flat (CF) approximation to
general relativity and the GW signal from the merger simulation is 
extracted up to quadrupole order.  
\end{abstract}
\section{Introduction}
Binary neutron star mergers belong to the strongest gravitational wave
sources for interferometer-type gravitational wave observatories as
LIGO \citep{ligo}, VIRGO \citep{virgo}, GEO600 \citep{geo600} and TAMA
\citep{ando}. After a long inspiral process which
lasts millions of years, the final merger phase takes place 
on a millisecond timescale.  The onset of merger is given when the two
companions become dynamically unstable near the innermost stable
circular orbit (ISCO) and mass transfer starts. Since the process has
no definite symmetries, 3D hydrodynamic 
simulations are necessary.  
In addition, because of the compactness of the system,
general relativistic effects have to be considered and treated as 
accurately as possible.

It has been pointed out that the gravitational wave frequency $f_{GW}$ 
relates to the compactness $(M/R)_\infty$ of a neutron star as 
$f_{\rm GW}\sim (M/R)^{3/2}_\infty$ near the ISCO, meaning that 
the gravitational wave frequency just before merger 
carries information on the radius of the neutron star. Hence the GW signal 
may constrain the equation of state (EoS) of high density matter,
\citep[see e.g.][]{lw96,rasiorev}. It has also been found recently 
that a neutron star formed after the merger 
may be supported by rapid and differential rotation 
even if its mass exceeds 60\% of the maximum mass of a single 
non-rotating neutron star, and the GW are emitted due to 
non-axisymmetric and quasiradial oscillations of the remnant for longer 
than the dynamical timescale \citep{shibataPRD,shibataPTP,stu03}.  
If these oscillations persist for a longer time, the integrated gravitational
wave may be detectable, carrying information on the high density matter. 

In addition to general relativistic gravity, 
various physics should be taken into account for many aspects of the binary 
neutron star merger problem such are the nuclear forces summarized in the EoS, neutrino physics and magnetic fields.  
So far, however, due to this complexity, investigations have concentrated 
either on the relativistic aspects of the problem using a simple EoS, or on the microphysical aspects while treating gravity in a
Newtonian framework. The pioneering work was done by Oohara \&
Nakamura in \citep[][ and references therein]{oohara}. Relativistic
aspects have been considered using the post-Newtonian approximation \citep[][ and references therein]{ayal, faber}, the
conformally flat approximation (\citealt{mathews};
\citealt{oechslinPRD};
\citealt[][and references therein]{fgr04}) and a fully general
relativistic treatment \cite[][ and references therein]{shibataPRD,shibataPTP,stu03}. 
Microphysical
improvements have been applied by \citet[][ and references
therein]{ruffert} and \citet[][ and references therein]{rosswog}
using different EoSs \citep{shen, lattimer} and a leakage scheme to
account for neutrino emission after the merger. For a review on the topic, 
see \citep{rasiorev}.

In this paper, we focus on two EoS for high density neutron star matter, one
describing pure nuclear matter, the other nuclear matter with a
transition to a quark phase at very high densities. A reason for 
investigating these EoSs 
is that the quark phase transition may be one of the most dramatic phenomena 
to change the compactness of neutron stars, and therefore 
its influence may be clearly observed in the GW 
spectrum of an inspiralling binary system.  
Following our previous work \citep[][]{oechslinPRD}, we consider the
merger problem in the conformally flat approximation. 

The paper is organised as follows.
In sect. \ref{sect:eqns}, we summarize the formalism to solve the
Einstein and the relativistic hydrodynamic equations. and describe the
numerical implementation, the choice of EoS and the initial conditions. In
sect. \ref{sect:results}, we present the results of the paper and
finally, we draw conclusions in sect. \ref{sect:conclusions}.

\section{Fundamental Ingredients}
\label{sect:eqns}

In this section we describe the numerical methods and the physics on
which our simulations are based. We consider a general relativistic
fluid whose internal properties are described by a given EoS. Two sets
of equations have to be solved simultaneously. On one hand side, we
consider the relativistic hydrodynamic equations governing the fluid
motion on the other side the Einstein equation of general
relativity determining the spacetime metric and therefore the
gravitational interaction. On top of it, an EoS which
closes the system of hydrodynamic equations, has to given as an input.

\subsection{Relativistic Hydrodynamics and Approximation 
for General Relativistic Gravity}

In the 3+1 decomposition of spacetime, the metric 
$ds^2=g_{\mu\nu}dx^\mu dx^\nu$can be written as 
\begin{equation}
ds^2=(-\alpha^2+\beta_i\beta^i)dt^2+2\beta_i dx^i dt+\gamma_{ij}dx^i dx^j,
\end{equation}
where $\alpha$ is the lapse function, $\beta^i$ is the shift vector,
and $\gamma_{ij}$ is the spatial metric. 
Spacetime quantities are decomposed with respect to the foliation using 
the hypersurface normal defined as 
$n_\mu = -\alpha\partial_\mu t$ with $n^\mu n_\mu=-1$ and 
a projection tensor defined as $\gamma_{\mu\nu}=g_{\mu\nu}+n_\mu n_\nu$, 
whose spatial component agrees with the spatial metric.  

The Einstein field equations can
be written as a set of two evolution equations
\begin{eqnarray}
\pa{t}\gamma_{ij}&=&-2\alpha K_{ij}+\nabla_{i}\beta_{j}+\nabla_{j}\beta_{i},\label{eqn:gammaevol}\\
\pa{t}K_{ij}&=&\alpha[R_{ij}-2K_{il}K^l_j+KK_{ij}\nonumber\\
&&-8\pi S_{ij}+4\pi \gamma_{ij}(S-\rho_E)]\nonumber\\
&&-\nabla_i\nabla_j\alpha+\beta^l\nabla_lK_{ij}+K_{il}\nabla_j\beta^l+K_{jl}\nabla_i\beta^l,\label{eqn:kijevol}\nonumber\\
\end{eqnarray}
and two constraint equations
\begin{eqnarray}
R+K^2-K_{ij}K^{ij}&=&16\pi \rho_E,\\
\nabla_{j}(K^{ij}-\gamma^{ij}K)&=&8\pi j^i
\end{eqnarray}
for the dynamic variables $\gamma_{ij}$ and $K_{ij}$, the
extrinsic curvature of the hypersurface \citep[see e.g.][]{numrel}.
Here, $R_{ij}$ is the Ricci tensor and $\nabla$ the covariant derivative 
associated with $\gamma_{ij}$.  The stress energy tensor 
$T_{\mu\nu}$ is decomposed into $\rho_E=n^\mu n^\nu T_{\mu\nu}$, 
the matter energy density, 
$j^i=\gamma^i_\mu n_\nu T^{\mu\nu}$, the matter momentum density
and $S_{ij}=\gamma_{i\mu} \gamma_{j\nu} T^{\mu\nu}$, 
the spatial projection of the stress energy tensor. 
Finally, $R, S$ and $K$ denote the traces of $R_{ij}$, $S_{ij}$ 
and $K_{ij}$, respectively.  

In the following discussion, we consider a perfect fluid with 
a matter stress energy tensor
\begin{equation}
T_{\mu\nu}=\rho h u_{\mu}u_{\nu}+pg_{\mu\nu}
\end{equation}
Here, $\rho$ refers to the rest mass density, $h=1+p/\rho+\epsilon$ to
the specific relativistic enthalpy, $\epsilon$ to the specific
internal energy, $u_{\mu}$ to the four velocity and $p$ to the fluid
pressure. Then,
\begin{eqnarray}
\rho_{E}&=&\rho h(\alpha u^0)^2-p\\
j^i&=&\rho h\alpha u^0 u^\mu \gamma_\mu^i.
\end{eqnarray}
The Lorentz factor $W=\alpha u^0$ can be calculated using
the normalisation condition $u_{\mu}u^{\mu}=-1$.
\begin{equation}
\alpha u^0 = (1 + \gamma^{ij}u_i u_j)^{1/2}.\\
\end{equation}

Isenberg proposed a waveless approximation to general relativity, 
in which he truncates some terms in Einstein's equation written 
in the ADM formalism 
to deduce an elliptic type formalism \citep[][]{Isenberg}.  
Later, the same set of equations have been rediscovered by Wilson and
Mathews \citep[][]{Wilson} 
and widely used to solve single or binary neutron stars problems.
\citep[e.g.][]{oechslinPRD} 
as well as binary black hole systems 
\citep[][]{ggb02}.  
The Isenberg-Wilson-Mathews theory has its own 
Hamiltonian as pointed out by Isenberg himself \citep[also discussed
in ][]{fus02}.  

In this approximation, the conformally flat condition for the spatial 
geometry $\gamma_{ij}=\psi^4\delta_{ij}$
and $K = \pa{t} K = 0$ are imposed, where $\delta_{ij}$ is the 
flat 3-metric. 
It may be viewed that the second one is a choice of 
a temporal gauge condition, the maximal slicing condition, while the first one is an approximation as well as partly a spatial gauge choice 
\citep[details can be found in][]{baumgarte}.
The remaining metric variables $(\psi,\alpha,\beta^i)$ do not satisfy all 
components of the Einstein's field equation consistently.  
We pick the constraints and the trace of the evolution equation for 
$K_{ij}$ as equations for the above five variables, 
which leads to the same set of equations derived from Isenberg's Hamiltonian mentioned above.   

The approximation leads to a considerable simplification of the
Einstein equations, since all metric equations can be written in elliptic form.

The trace of the evolution equation for $K_{ij}$
together with the maximal slicing condition $K=0$ leads to an 
equation for the lapse
\begin{equation}
\Delta(\alpha\psi)=2\pi\alpha\psi^5(\rho_E+2S)+\frac{7}{8}\alpha\psi^5 K_{ij}K^{ij},
\label{eqn:alphapsi}
\end{equation}

The trace-free part of the evolution equation for $\gamma_{ij}$
together with the conformally flat condition $\gamma_{ij}=\psi^4\delta_{ij}$ provides
\begin{equation}
\label{kijform}
2\alpha\psi^{-4}K_{ij}=\partial_j\tilde\beta_i+\partial_i\tilde\beta_j
-\frac{2}{3}\delta_{ij}\partial_k\beta^k, 
\end{equation}
where $\partial_i$ is a derivative with respect to the coordinate 
associated with $\delta_{ij}$ and 
$\tilde \beta_i$ relate to the shift $\beta^i$ through 
the flat metric as $\tilde\beta_i = \delta_{ij}\beta^j$.  
The Hamiltonian constraint provides an equation for the conformal factor
\begin{equation}
\Delta\psi=-2\pi\psi^5 \rho_E-\frac{1}{8}\psi^5 K_{ij}K^{ij}\equiv 4\pi S_{\psi}.
\label{eqn:psi}
\end{equation}
Finally, with the momentum constraint we obtain an equation for 
the shift vector
\begin{eqnarray}
\Delta\beta^i+\frac{1}{3}\pau{i}\pa{j}\beta^j
&=&\pa{j}\ln\left(\frac{\alpha}{\psi^6}\right)
\left(\pau{j}\beta^i+\pau{i}\beta^j
-\frac{2}{3}\delta^{ij}\pa{l}\beta^l\right)\nonumber\\
&+&16\pi\alpha \psi^4 j^i, 
\label{eqn:momentumc}
\end{eqnarray}
where $\pau{i} = \delta^{ij}\pa{j}$. 
Using the definition
\begin{equation}
\beta^i=B^i-\frac{1}{4}\pau{i}\chi,
\end{equation}
(\ref{eqn:momentumc}) splits into two simpler parts
\begin{equation}
\Delta B^i=\pa{j}\ln\left(\frac{\alpha}{\psi^6}\right)
\left(\pau{j}\beta^i+\pau{i}\beta^j-\frac{2}{3}\delta^{ij}\pa{l}\beta^l\right)
+16\pi\alpha \psi^4 j^i,
\label{eqn:b}
\end{equation}
and
\begin{equation}
\Delta\chi=\pa{j}B^j.
\label{eqn:chi}
\end{equation}
While the trace-free components of the evolution-equation for $\gamma_{ij}$
(\ref{eqn:gammaevol}) are used to relate $K_{ij}$ and the metric
variables as in equation (\ref{kijform}), its trace part 
and the trace-free components of the evolution-equation
for $K_{ij}$ (\ref{eqn:kijevol}) are dropped by the CF
approximation. The trace of (\ref{eqn:gammaevol}) can be used to check
the accuracy and reliability of the CF approximation
\citep{dimmelmeier, oechslinDiss}

The relativistic hydrodynamic equations can be written in the
following way 
\begin{eqnarray}
{\partial_t \rho^*}+{\partial_i (\rho^* v^i) }&=&0,\\
{\partial_t (\rho^* \epsilon ) }+{\partial_i (\rho^* \epsilon v^i)}&=&
        -p\biggl[{\partial_t (\alpha u^0 \gamma^{1/2})}\nonumber\\
	&&+{\partial_i (\alpha u^0 \gamma^{1/2} v^i) 
	}\biggr],\\
{\partial_t (\rho^* h u_k)} +{\partial_i (\rho^* h u_k v^i) }
	&=&-{\alpha \gamma^{1/2}}{\partial_k p}+\\
	\rho^* h \biggl[-\alpha u^0 {\partial_k \alpha }
	&+&u_j {\partial_k \beta^j }
	-\frac{u_i u_j }{ 2 u^0} 
	{\partial_k \gamma^{ij} }\biggr],
\end{eqnarray}
where
\begin{eqnarray}
\rho^*&=&\rho\alpha u^0\text{det(}\gamma_{ij}\text{)}\quad\text{and}\\ 
v^i&=&-\beta^i+\frac{\gamma^{ij} u_j}{u^0}
\end{eqnarray}
In the case of conformal flatness, the system reduces to
\begin{eqnarray}
\frac{d}{dt}\rho^*&=&-\rho^*\partial_i v^i\label{conteqn}\\
\frac{d}{dt}\tilde{u}_i&=&-\frac{1}{\rho^*}\alpha\psi^6\partial_{i}
p-\alpha\tilde{u}^0\partial_{i}\alpha+\tilde{u}_j\partial_{i}\beta^j\nonumber\\
&&+\frac{2\tilde{u}_k\tilde{u}_k}{\psi^5\tilde{u}^0}\partial_{i}\psi\\\label{momentumeqn}
\frac{d}{dt}\epsilon&=&-\frac{p}{\rst}\partial_{i}v^i-\frac{p}{\rst}\partial_t\ln(\alpha u^0\psi^6),\label{energyeqn}
\end{eqnarray}
with
\begin{eqnarray}
\tilde{u}_i&=&hu_i,\\
\rst&=&\rho\alpha u^0\psi^6\qquad\text{and}\\
v^i&=&-\beta^i+\frac{\delta^{ij} u_j}{\psi^4 u^0}
\end{eqnarray}
Here, we have cast the system into a Lagrangian formulation
which is better suited for an implementation on a computer using a
Lagrangian scheme like SPH (cf.
sect. \ref{sect:implementation}). The conserved hydrodynamic variables
are $(\rst, \tilde{u}_i, \epsilon)$ whereas the primitive variables
are $(\rho, v^i, \epsilon)$.

\subsection{Numerical implementation}
\label{sect:implementation}

To solve the system of hydrodynamic equations (\ref{conteqn}) -
(\ref{energyeqn}) we use the Smoothed Particle Hydrodynamics method (SPH;
\citep{benz, monaghan}), which is widely used in
astronomical and astrophysical simulations. The original form of SPH
that solves the Newtonian hydrodynamic equations is presented in
\citep{benz}. We list here the generalized version which is
appropriate for the above relativistic hydrodynamic equations. For
details, see \citep{oechslinPRD}. The continuity
equation turns into a relation between $\rst$ and the rest masses
of the individual particles
\begin{equation}
\rho _{a}^{*}=\sum _{b}m_{b}W_{ab},
\end{equation} 
where $m_{b}$ is the rest mass of particle $b$ and
$W_{ab}=W(|\mathbf{r}_{a}-\mathbf{r}_{b}|,h_{SPH})$ denotes the weight given
by the standard spherical spline Kernel function
$W(\mathbf{r},h_{SPH})$. The rest masses $m_b$ and the smoothing
length $h_{SPH}$ are initially chosen to fit
an initial spatial distribution of $\rst$. The pressure gradient
contained in the momentum equation (\ref{momentumeqn}) is calculated
as in standard SPH but replacing $\rho$ by $\rho ^{*}$,
\begin{equation}
\frac{1}{\rho _{a}^{*}}\pa{i} p_{a}=-\sum _{b}m_{b}\left( \frac{p_{b}}{\rho _{b}^{*2}}+\frac{p_{a}}{\rho ^{*2}_{a}}\right) \pa{i} W_{ab}.
\end{equation}
The energy equation (\ref{energyeqn}) contains a pdV-term similar to Newtonian SPH
\begin{equation}
\left. -\frac{p}{\rho}\pa{i} v^{i}\right| _{a}=\frac{1}{2}\sum _{b}m_{b}\left( \frac{p_{a}}{\rho ^{*}_{a}\rho _{a}}+\frac{p_{b}}{\rho ^{*}_{b}\rho _{b}}\right) ({v}^i_{a}-{v}^i_{b})\pa{i} W_{ab}.
\end{equation}
The additional terms in the momentum and energy equation that arise
from the gravitational interaction are evaluated by first solving the
equations for the metric variables $\alpha$,$\psi$ and $\beta^i$ on a
overlaid grid (see discussion below) and by mapping back those
quantities onto the particles by second-order interpolation. The total
time derivative of $\ln (\alpha u^{0}\psi ^{6})_{a}$ in the energy
equation is evaluated using second order finite differencing in time.

On top of these equations, an artificial viscosity (AV) scheme is
implemented with time-dependent viscosity parameters \citep{morris97}
which have only significant values in the presence of shocks. The AV
produces an additional viscous pressure which is added to the physical
fluid pressure \citep{siegler97}.

The field equations (\ref{eqn:alphapsi},\ref{eqn:psi},\ref{eqn:b} and
\ref{eqn:chi}) are discretized on a grid which covers the matter
distribution. The evaluation proceeds generally in three steps
\begin{itemize}
\item First, we calculate the source corresponding to the potential.
The involved hydrodynamic quantities defined on the SPH particles
are transferred onto the grid by assigning SPH-interpolated
values to the grid points, i.e.
\begin{equation}
\langle f(x)\rangle_{SPH}=\sum_j f_j\frac{m_j}{\rho^*_j}W(|\bmath{r}-\mathbf{r}_j|,h_j).
\end{equation}
\item Afterwards, the potential is obtained by solving the
corresponding Poisson-equation. The $\psi$- and the
$\alpha\psi$-equation are closely linked together and are therefore
solved in two iteration steps using the solution from the previous
timestep as a guess. The remaining four equations determining the shift
vector are also closely coupled and are solved simultaneously within
the FMG algorithm. 
\item All needed derivatives are calculated on the grid using finite differencing.
\item Finally, the potential and its derivatives are mapped back onto the
SPH particle distribution using a triangular shaped cloud (TSC) method \citep{hockney} 
known from particle mesh codes. This is equivalent to a second order interpolation.
\end{itemize}

Boundary conditions and extensions of the solution beyond the grid are
obtained using a multipole expansion of the source term. For details we refer to 
\citep{oechslinPRD}.

Like the Newtonian and the first-order post-Newtonian (1PN)
approximation, the CF approximation does not include gravitational
radiation and its reaction by construction. We therefore have to
add an additional gravitational wave (GW) extraction scheme and a radiation
reaction force which accounts for the angular momentum and energy
carried away by GW. The waveform in transverse-traceless gauge is
extracted in the slow-motion limit \citep{thorne80, Wilson} and up to quadrupole order using
\begin{equation}
h^{TT}_{ij}(t,\mathbf{r})=\frac{2}{r}P_{ijkl}\ddot Q_{kl}(t-r),
\end{equation}
where
\begin{equation}
Q_{ij}=\text{STF}\left\{-2\int S_{\psi}(x)x_i x_j d^3x\right\}
\end{equation}
is the mass-quadrupole of the system. The radiation-reaction force is chosen in a similar way to the original
Burke-Thorne expression $F^{rr}_i=\sigma\pa{i}V$ with
$V=-1/5x_ix_jQ_{ij}^{(5)}$, but its strength $\sigma$ is chosen such 
that the resulting angular momentum loss by backreaction
$\dot{J_i}^{tot}-\dot{J_i}^{num}$ reproduces the expression
\begin{equation}
\dot{J_i}=-\frac{2}{5}\epsilon_{ijk}Q_{jm}^{(2)}Q_{km}^{(3)}
\label{eqn:quad}
\end{equation}
of the slow motion GW extraction approximation up to a difference of
the order of $\simeq 5\%$.  
Since $J_i=\epsilon_{ijk}\int \rst (x^j\tilde
u_k-x^k\tilde u_j)\, d^3x$, we can obtain the angular momentum change using 
\begin{equation}
\dot{J_i}=\epsilon_{ijk}\int \rst \left((x^j
\dot {\tilde u}_k-x^k\dot{\tilde u}_j)+(v^j\tilde u_k-v^k\tilde u_j)\right)d^3x\label{eqn:dotJ}.
\end{equation}
The expressions $\dot{J_i}^{num}$, the numerical angular momentum
error, and $\dot{J_i}^{tot}$, the total angular momentum change per
unit time, are obtained by evaluating (\ref{eqn:dotJ}) before adding and
after adding the backreaction force terms, respectively. 
Since $\dot{J_i}^{num}$ has an oscillating behaviour around zero, 
as a result thereof, $\dot{J_i}^{tot}$ depends on the binary orbit.  
Therefore, in practical determination of the value of $\sigma$, 
we take an average $\langle\dot{J_i}^{tot}-\dot{J_i}^{num}\rangle$ 
for about a half orbit, and compare with 
$\langle \dot J \rangle$ computed from the quadrupole 
formula (\ref{eqn:quad}) that is averaged for the same orbit.  
For model B1 we obtain for the ratio 
$\langle\dot{J_i}^{num}\rangle/\langle\dot{J_i}^{tot}\rangle\sim0.3\%$ at
the ISCO, where the average is made over one orbital
period, while at a certain period of time $\dot{J_i}^{num}$ 
may become comparable to $\dot{J_i}^{tot}$.  
Since the ratio $\langle\dot{J_i}^{num}\rangle/\langle\dot{J_i}^{tot}\rangle$
is negligible, the radiation reaction force is
clearly driving the inspiral process, although the predicted time 
to merger have to be
taken with care due to the approximative radiation reaction
scheme and the spatially conformal flat assumption used in 
our initial conditions and simulations.  
Fully relativistic inspiral simulations \cite{miller2} 
are intrinsically consistent concerning radiation reaction 
dissipation but they still depend on the outer 
boundary location.

\subsection{Equation of state}

To investigate the influence of quark matter to a neutron star merger,
we consider two equations of state, an EoS describing pure nuclear
matter (``hadronic EoS'') and an EoS describing nuclear matter with a
phase transition to quark matter at very high densities (``hybrid
EoS'').

The hadronic EoS is realized using the
non-linear $\sigma$-$\omega$-model in the relativistic mean field
approximation with the TM1 parameter set which is
motivated by a least-square fit to experimental results including
stable and unstable nuclei \citep{sugahara}. At densities above $\rho\gtrsim
10^{14}\text{g/cm}^3$ this is a good approximation. For lower values,
when inhomogeneous nuclear matter appears and the non-linear
$\sigma$-$\omega$ model is not valid any more, other approximations have to be
used \citep[e.g. ][]{shen}. In our case, we append a polytropic EoS
$p=\kappa\rho^\Gamma$ where $\kappa$ and $\Gamma$ are adjusted to
fulfill smoothness of pressure and internal energy. We choose the
transition density to the polytropic EoS at
$2\times10^{14}\text{g/cm}^3$ which leads to $\Gamma\simeq2.86$, similar
to common realistic EoSs in this density regime.
The hybrid EoS is obtained by combining the hadronic EoS with a MIT 
bag model \citep{MITmodel} using a variable pressure
phase transition construction \citep{glendenning}. We use massless up
and down quarks and a Bag constant of $90\text{MeV/fm}^3$. The resulting 
hybrid EoS then describes three physical phases
\begin{itemize}
\item A pure hadronic phase below 
$\approx5\times10^{14}\text{g/cm}^3\approx1.8\rho_0$ 
where the EoS coincides with the hadronic EoS. Here,
$\rho_0:= 2.8\times10^{14}\text{g/cm}^3$ is the nuclear saturation
density. The stiffness in this
region varies between $\Gamma\approx3$ and $\Gamma\approx2.5$.
\item A quark-hadron mixed phase between 
$\approx5\times10^{14}\text{g/cm}^3$ and $\approx10^{15}\text{g/cm}^3$ 
$\approx3.5\rho_0$ where both quark and hadrons are present. 
In this phase transition region, the EoS substantially softens with $\Gamma\approx 1-1.5$.
\item A pure quark phase above $\approx10^{15}\text{g/cm}^3$ where the
MIT-bag-model EoS with similar adiabatic index but lower absolute 
pressure in comparison to the hadronic EoS is applied.
\end{itemize} 

\begin{figure}
\begin{center}
\vspace{1cm}
\includegraphics[width=8cm]{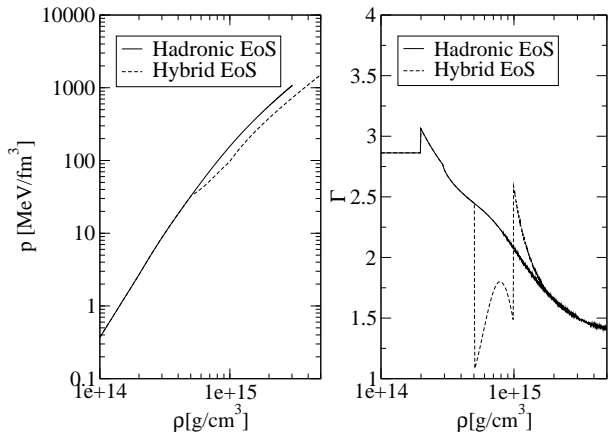}
\caption{Comparison of the considered nuclear EoS. The hybrid EoS
(dashed curve) has a phase transition region ($1.8\rho_0<\rho
<3.5\rho_0$) where adiabatic indices are substantially lower, followed
by a quark phase which is has a similar stiffness as the nuclear
matter at those densities.}
\label{fig:nuceos_gamma}
\end{center}
\end{figure}

In both EoSs, we neglect temperature effects, i.e. we set T=0. The
redundant internal energy information from the EoS is dropped and we
consider the pressure as a function of the density alone. This is a
good approximation in the high-density regime where thermal effects to
the pressure
are rather small. At lower densities (e.g. in the disk around the
merger remnant) the thermal component is certainly not
negligible. However we concentrate in this work on the high-density
merger remnant so we claim to be rather accurate with this
approximation although thermal effects could have an effect via
neutrino losses.

\subsection{Initial Conditions}
To construct initial conditions, we assume the binary to be in a
quasi-equilibrium state. This assumption is not very well satisfied in
the very vicinity of the ISCO \citep{miller} and we therefore choose
slightly wider initial orbits. In our models the binaries have
initially no radial velocity either and need about half an
orbit to begin with the inspiral.
We apply the method of Ury\=u \& Eriguchi (2000) to construct an initial
configuration. This method takes an EoS
as an input and then solves the hydrostatic equation for an irrotational 
velocity field together with the Einstein
field equations in the CF approximation. A similar method has been
developed by \citep{bonazzola}. We map the output of the above method,
the hydrodynamic quantities and the conformal factor onto a distribution of SPH
particles. Finally, this SPH distribution is relaxed to avoid spurious
inter-particle forces using a braking term
\begin{equation}
\mathbf{f}=-\frac{1}{\tau_{relax}}(\mathbf{v}-\mathbf{v}_{irr}),
\end{equation}
where $\mathbf{v}_{irr}$ is the given initial irrotational velocity field
of the initial mass distribution.
Using this method, we produce a set of irrotational quasi-equilibrium
configurations, both varying the EoS and the gravitational mass
$M_\infty$, where $M_\infty$ refers to the mass of a single NS in isolation. 
For $M_\infty$, we use values between $M_\infty=1.2M_\odot$ which is at the 
lower limit of the observational range
and $M_\infty=1.75M_\odot$ which is close to the maximal gravitational NS mass
of $M_{max}=1.78M_\odot$ of the hybrid EoS. The
models are summarized in Tab. \ref{tab:initconf}. The letter part
(A-E) in the model label indicates the mass while the number part
(1,2) indicates the EoS. All orbits are taken slightly outside the
ISCO, about $12\sim 25\%$ in coordinate separation. This results in
smaller lag angles and its oscillations in the initial phase since tidal forces are
weaker at larger separation. In Tab. \ref{tab:resconf}, we summarize
properties which result as a consequence of the chosen initial
parameters in Tab. \ref{tab:initconf}.

\begin{center}
\begin{table}
\caption{Input parameters of the considered models using the
hadronic and hybrid EoS. $M_\infty$ denotes the gravitational mass of
one NS in isolation and $C$ is the compactness. Both the masses and the compactness refer 
to one binary component in isolation. $N_{Particle}$
denotes the number of SPH particles and $N_{Grid}$ is the number of grid
points. In model E1, $97^3$ gridpoints are used, but at
the onset of collapse, the resolution is enhanced to $129^3$. Models
with a '1' in the label are calculated with the hadronic EoS, 
models with a '2' used the hybrid EoS.}
\label{tab:initconf}
\begin{tabular}{l c c c c}
\hline
Model & $M_\infty[M{}_\odot]$ & $C=(M/R)_\infty$ &$N_{particle}$&$N_{Grid}$\\
\hline
A1 & 1.2 &0.1276&154038&$97^3$\\
B1,B2 & 1.35 &0.1424&153993&$97^3$\\
C1,C2 & 1.4 &  0.1475 &169798 &$97^3$\\
D1 & 1.5 &0.1577&100488&$97^3$\\
D2 & 1.5 &0.1579&100548&$97^3$\\
E1 & 1.75 &0.1853&162536&$97^3 (129^3)$\\
E2 & 1.75 &0.1959&157222&$129^3$\\
\end{tabular}
\end{table}
\end{center}

\begin{table}
\caption{Selected properties of our initial models resulting from the
input parameters shown in Table \ref{tab:initconf}. $M_0$ is the
restmass of one single NS, $P$ the orbital period, $f_{GW,0}$ the
corresponding GW frequency, $d_0$ the initial orbital separation and
$R_\infty$ the radius of one single isolated NS measured in
Schwarzschild coordinates. Note that the GW frequencies are lower than those in Fig. \ref{fig:ISCOfreq} which are
taken at the ISCO.}
\label{tab:resconf}
\begin{tabular}{l c c c c c}
\hline
Model &  $M_0[M{}_\odot]$ & $P [ms]$ & $f_{GW,0}=2/P$ & $d_0 [km]$ & $R_\infty [km]$\\
\hline
A1 &1.295 & 2.92 & 685 & 38.00 & 13.88\\
B1,B2 & 1.471 & 2.72 & 737 & 36.93 & 13.99\\ 
C1,C2 &1.523& 2.65 & 756 & 36.71 & 14.01\\
D1 &1.649& 2.57 & 777 & 36.19 & 14.03\\
D2 &1.649& 2.57 & 777 & 36.19 & 14.02\\
E1 &1.957& 2.20 & 909 & 33.40 & 13.93 \\
E2 &1.959& 1.97 & 1058 & 31.14 & 13.07\\
\end{tabular}
\end{table}

\section{Results}
\label{sect:results}

In this section, we consider the final evolution of the binary NS from 
a circular orbit configuration to one single object, a transient 
NS or a BH. This evolution takes place on a dynamical timescale and 
is triggered by radiation reaction.  
While in Sec.~\ref{sect:quasiequilibrium}, the EoS effects on the GW
frequency from the final inspiral phase just before a merger 
has been discussed, in the latter subsections the result of the 
dynamical evolution of the binary NS from stable circular orbit to 
the merger is presented.  

\subsection{EoS impact on the quasi-equilibrium binary}
\label{sect:quasiequilibrium}
By considering the compactness $(M/R)_\infty$ in Tab.\ref{tab:initconf}, 
we observe that the properties of the models become EoS dependent for the
most massive cases D1, D2 and E1, E2, i.e. when $M_\infty \gtrsim
1.5M_\odot$. For lower stellar masses, the NSs do not reach the phase
transition density $\rho_t := 1.8\rho_0$ in the center and an EoS effect cannot be observed. 
It has been shown that the orbital frequency and the gravitational wave frequency of a binary system depends on the
compactness of its components \citep[e.g.][]{lw96, uryu104015, faberletter}. 
Therefore we expect that an EoS effect may be seen in the GW frequency 
at the final stage of inspiral in our models.  

\begin{figure}
\includegraphics[width=8cm]{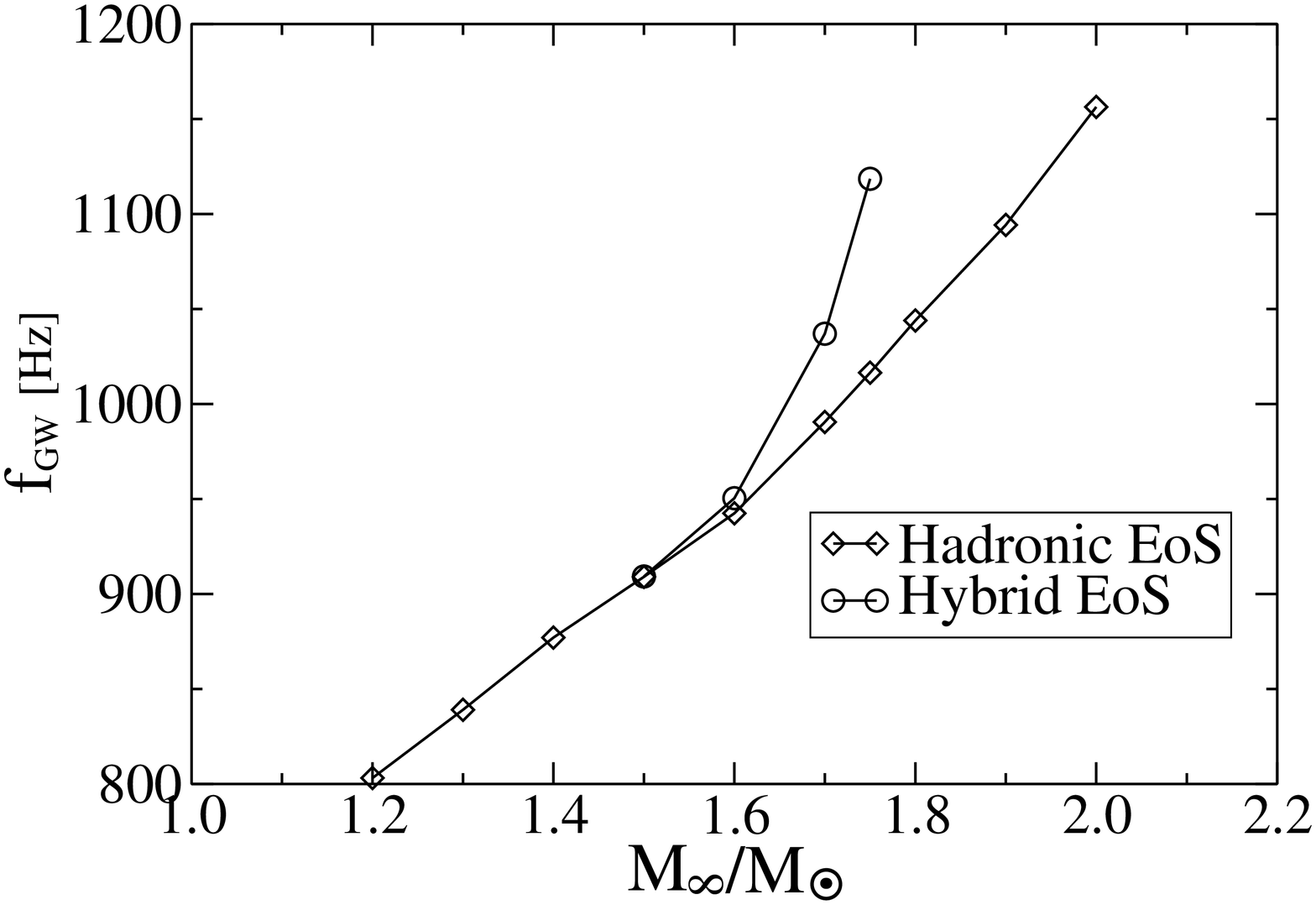}
\includegraphics[width=8cm]{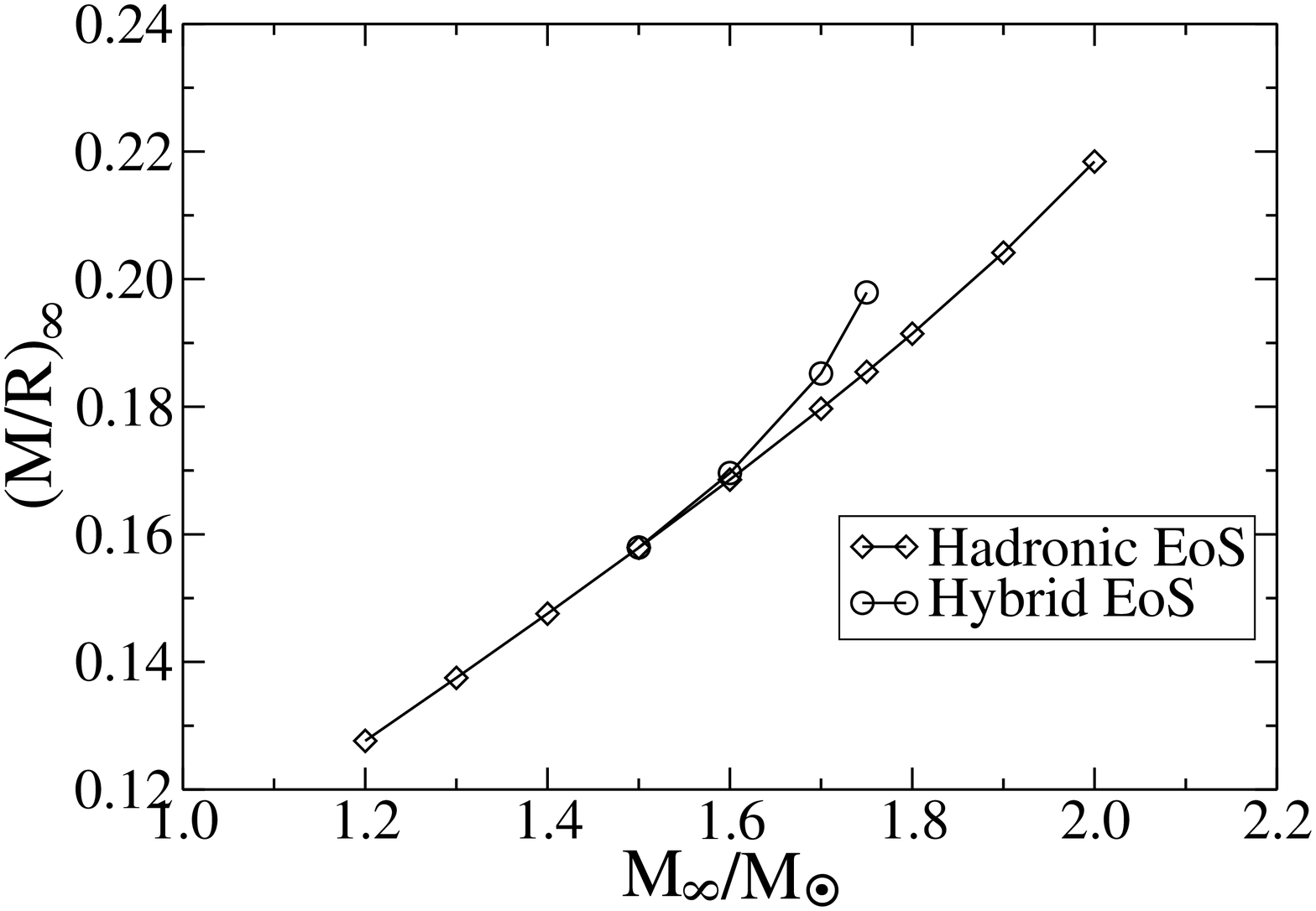}
\caption{(a)GW frequencies depending on the gravitational mass in
isolation at the ISCO.
(b)Compactness of an isolated neutron star 
with respect to the gravitational mass.}
\label{fig:ISCOfreq}
\end{figure}

Using the method of \citep{uryumethod,uryu104015}, we systematically 
investigated the locations of the ISCO, beyond which the binary become 
dynamically unstable, and the GW frequency at the ISCO.  As shown in 
Fig.~\ref{fig:ISCOfreq}(a), an EoS effect becomes important above 
$M_\infty=1.5M_\odot$.  The reason is that at this mass, the central 
density reaches the phase transition density to mixed matter in the hybrid 
case and a dense core of mixed matter is formed. For even larger masses, 
the transition density to pure quark matter at $\rho\simeq3.5\rho_0$
is reached and a quark core is formed.  Due to this, a neutron star with the hybrid EoS becomes more 
compact than one with the hadronic EoS as shown in Fig.~\ref{fig:ISCOfreq}(b), 
which is reflected in the difference in the GW frequency.  Near the maximal 
NS mass of the hybrid EoS, the GW frequency of the hybrid EoS binary is up 
to 10\% larger than that of the hadronic EoS binary.  Such a change of 
tendency for increasing gravitational wave frequency with respect to 
increasing mass at $M_\infty \agt 1.5M_\odot$ may suggest an existence
of a drastic change of EoS such as a quark phase transition.  

\subsection{Dynamical Evolution of the Neutron Star Merger}
\label{sect:evol}
\begin{figure*}
\begin{minipage}[t]{17cm}
\includegraphics[width=17cm]{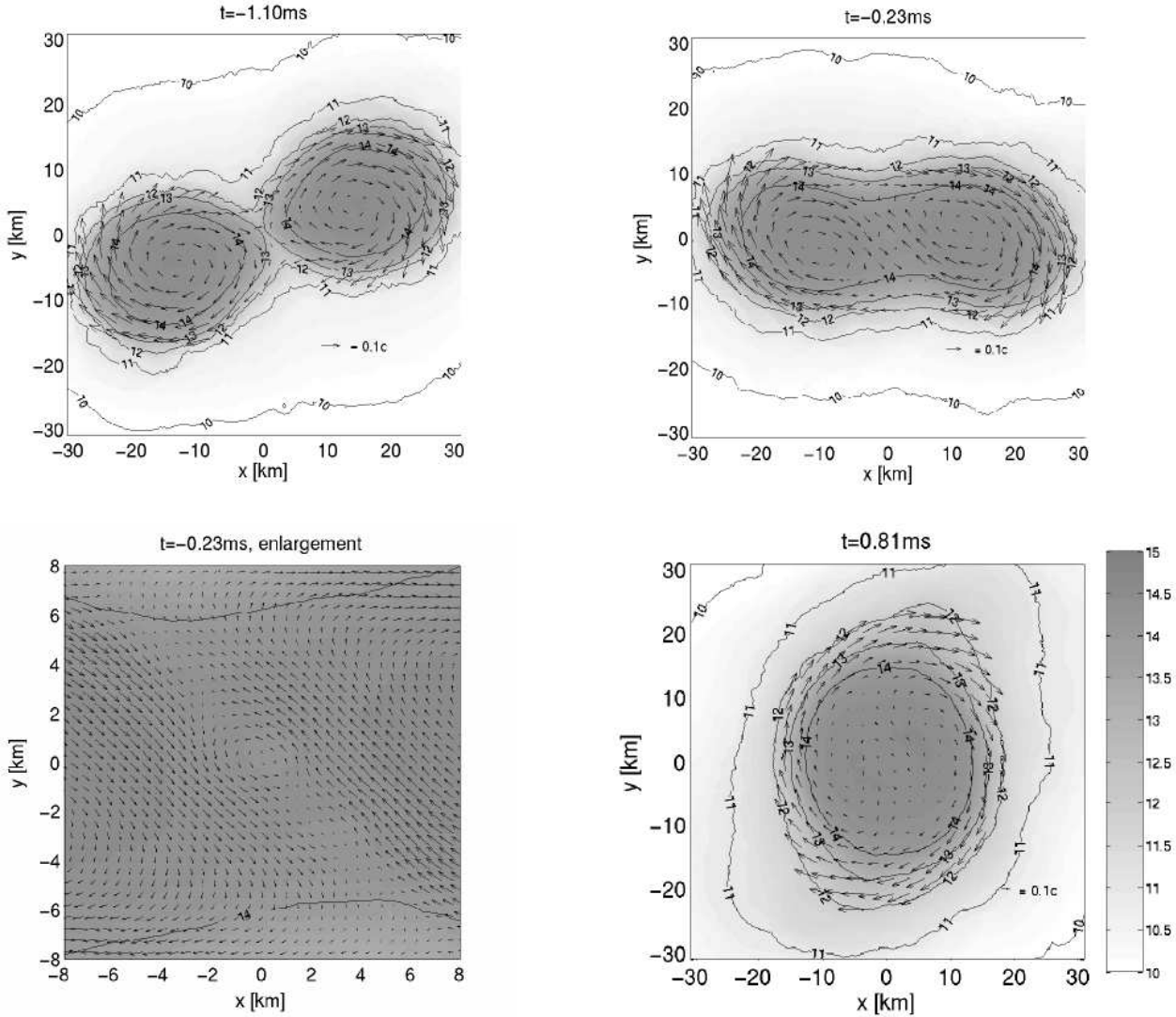}
\caption{Snapshots at selected times of the baryonic density
distribution and the velocity field of model B1 during pre-merger and
merger evolution. The density is plotted logarithmically in units of
g/cm${}^3$, the velocity field is plotted in the corotating frame.}
\label{premerger}
\end{minipage}
\end{figure*}

\begin{figure*}
\begin{minipage}[t]{17cm}
\includegraphics[width=17cm]{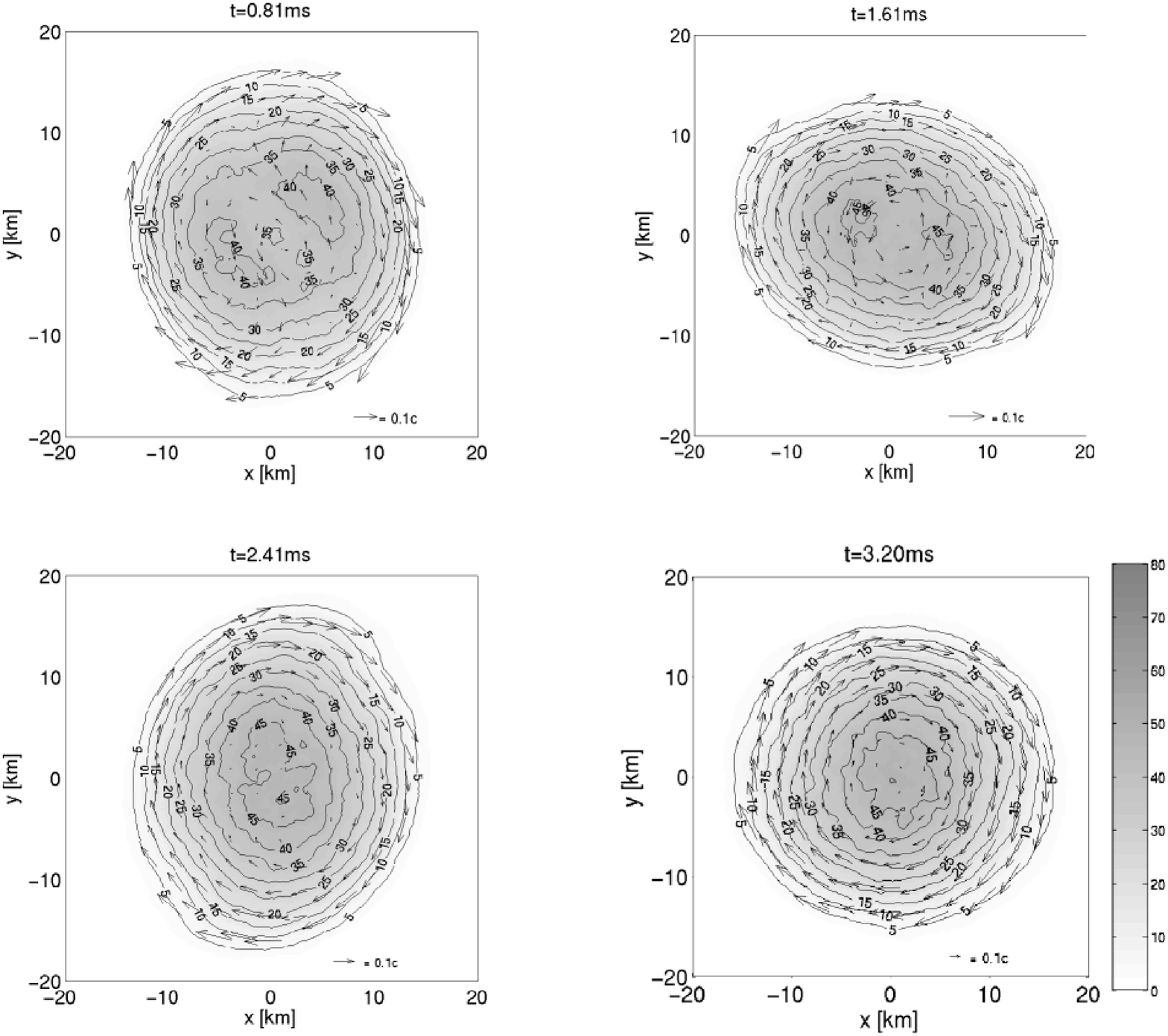}
\caption{Same as Fig.\ref{premerger} but for the post-merger
evolution. The density is plotted linearly in units of $10^{13}$g/cm${}^3$.}\label{postmerger}
\end{minipage}
\end{figure*}

We now turn to the merger phase that follows the quasi-stationary inspiral
when the two NS cross the ISCO and become dynamically unstable.
To first illustrate the overall dynamics, we plot in
Fig. \ref{premerger} the density distribution for model B1 together
with the velocity field in the corotating frame, whose angular 
velocity is determined as
\begin{equation}
\Omega=\frac{\sum_i m_i \omega_i}{\sum_i m_i}
\end{equation}
where $i$ runs over all particles and $\omega_i$
denotes the angular velocity of the individual
particles. The stars are counterrotating
relative to the corotating frame and tidal lag angles are
developing (snapshot 1). At merger, this counterrotation is first
turned into a shear motion
at the contact layer and then into a pattern of vortex rolls due to
the growing Kelvin-Helmholtz instability (snapshot 2,3). Finally we
end up with a differentially rotating, non-axisymmetric merger remnant
(snapshot 4). In Fig. \ref{postmerger}, we concentrate on the evolution
of the merger remnant during the first milliseconds by plotting the
density contour linearly which allows a better visualisation of the
high-density central parts (The lowest density contour is at $5 \times 10^{13}\text{g/cm}^3$.)  In this figure we can see that
during a phase of $\sim 2$ms after merger, the core actually consists
of two small subcores which are leftovers of the original NS
cores. Visible is also that the two subcores still carry a small
counterrotating motion - a direct consequence of the irrotating setup
- which disappears continuously until we end up in an nearly axisymmetric,
differentially rotating configuration. It is reported from
finite difference simulations \cite[e.g.][]{shibataPTP,stu03} that
this twin-core pattern persists for a longer time than our
simulations. This may be a consequence of the numerical viscosity in our code. 

\subsection{EoS differences in maximal density}
\label{sect:rhomax}
\begin{figure}
\includegraphics[height=21.5cm,width=8cm]{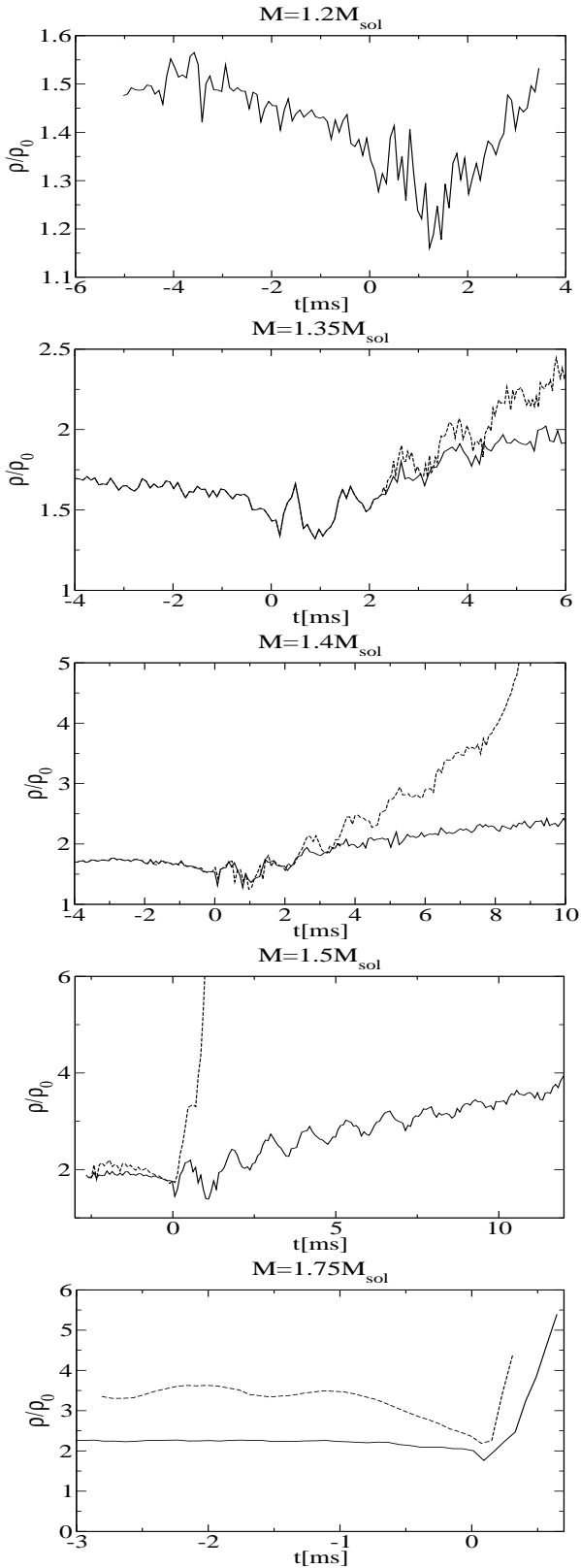}
\caption{Evolution of the maximal density measured in units of the nuclear
saturation density $\rho_0:=2.8\times10^{14}$g/cm${}^3$. The origin of the
time axis has been shifted to the GW luminosity maximum. Solid lines
correspond to hadronic models, dashed lines to hybrid models.}
\label{fig:rhomax}
\end{figure}

To investigate the different consequences of the EoSs in detail, we first
consider the evolution of the maximum density during the
merger phase. We split the whole merger event in a pre-merger evolution where the
binary still consists of two tidally stretched objects and in a following
post-merger evolution where a merger remnant, a NS or BH is forming.
Since the SPH particle density $\rho_a$ has no direct physical
meaning, the actual density $\rho(x)$ has to be calculated as a
statistical average over the particles densities $\rho_a$. To find the
maximal density, we first determine
density values on selected gridpoints and then look for their
maximum. Both the statistical averaging and the finite grid spacing
introduce a small noise. In general, the evolution of $\rho_{max}$, plotted in Fig. \ref{fig:rhomax}
for our simulations, shows a slow decrease during the pre-merger phase when
the two stars become tidally stretched and reaches its minimum
during the actual merger. This is the case when the GW luminosity either
approaches its maximum and the two NS are maximally tidally stretched
or just a bit later when a new merger remnant has already
formed but a twin-core structure is still present (see
sect. \ref{sect:evol}). After the minimum, we can
distinguish three possible evolutionary scenarios of $\rho_{max}$,
depending on the collapse behaviour of the remnant. If the remnant
can be stabilized by pressure and centrifugal forces, $\rho_{max}$
slowly increases and then becomes constant. The second scenario is the delayed
collapse where $\rho_{max}$ first slowly increases on several
dynamical timescales. This fragile equilibrium is concluded by a final
collapse, which is visible as a steep rise of $\rho_{max}$. The last scenario is the
immediate collapse just after the merger. Here, $\rho_{max}$ increases
without delay on a dynamical timescale.

Differences in the merger dynamics caused by the EoS,
appear when $\rho_{max}$ crosses the phase transition density
$\rho_t$. At this point the hybrid EoS and therefore the dynamical evolution of the hybrid models
separates from that of the hadronic EoS.
If the individual stars are more massive than $M_\infty\simeq1.5 M_\odot$,
their central density exceeds the phase transition density $\rho_t$ . In this
case, EoS differences happen already before merger. This makes a very
small effect which even disappears close to merger due to tidal
distortion if $M_\infty$ is just at the threshold of $\simeq1.5M_\odot$ as in model D1/D2. On the other hand, if $M_\infty$ is
close to the maximal gravitational mass of the hybrid EoS as in model
E1/E2, the EoS difference in $\rho_{max}$ increases to up to
$\sim60\%$. Differences are also measurable for global
quantities like the angular velocity of the binary, the compactness
and especially for the GW frequency (cf. sect. \ref{sect:quasiequilibrium}).
If the initial mass is less than $M_\infty\simeq1.5 M_\odot$ an EoS
difference can only be seen after the merger as part of a different
evolution scenario of the merger remnant. Typically, the remnant collapses immediately or after a
couple of dynamical timescales in the hybrid cases while it settles
down to a transient NS in the hadronic cases. Among the hadronic
models, there is only the very massive model E1 which collapses
immediately. In model D1, the maximal density slowly increases but
does not really settle down - a sign that this object might eventually
collapse. However, this might well be a numerical effect as
angular momentum is transported from the core to the outer layers and
differential rotation is slowly converted into uniform rotation. The
amount which is transported is on the order of 40-50\% in the very
center during the whole evolution, despite of a very good overall
angular momentum conservation, either by numerical viscosity or by
gravitational interaction between the non-axisymmetric core and the outer 
layers.  However it is still possible to stabilize a remnant with a
gravitational mass of nearly $3M_\odot$ with an EoS having a maximal gravitational mass of
$2.2M_\odot$ for about 12ms $ \simeq820M_{tot}$, where
$M_{tot}=2.98M_\odot$ is the initial gravitational mass
for this model. This stabilisation effect due to differential
rotation has been pointed out in \citep{shibataPRD,baumgartediffrot}. 
The less massive models C1,B1 and A1 lead to transient NS which 
do not collapse on a hydrodynamic
timescale. On the hybrid-EoS model side, the models collapse with different
 collapse timescales. While models E2 and D2 collapse
immediately, C2 collapses with a delay of about $\sim9$ms. Model B2
does not collapse within the simulation time but it also shows a
continuously growing maximum density. However, these collapse
timescales are likely to be dominated by the above mentioned angular
momentum transport.

\subsection {The Gravitational Wave signal}
\begin{figure}
\includegraphics[height=21.5cm,width=8cm]{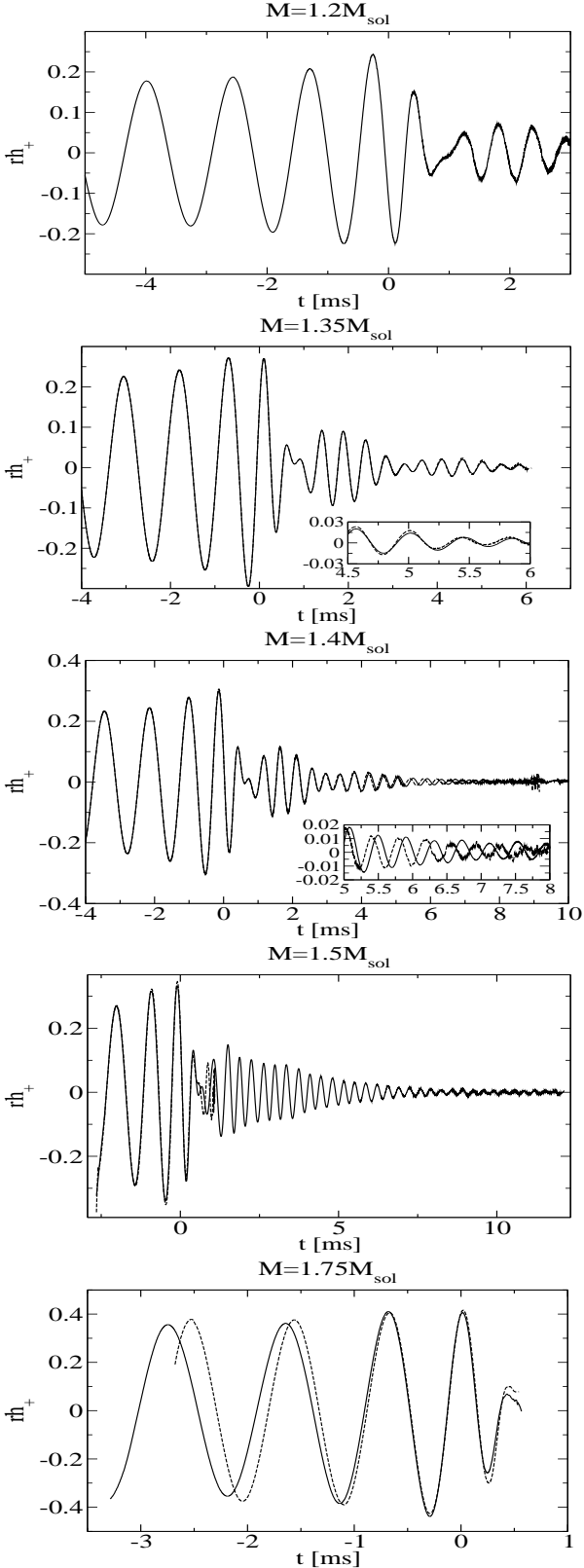}
\caption{Gravitational waveforms of all models sorted according to
their initial mass. The origin of the time axis has been shifted to
the GW luminosity maximum. Solid lines correspond to hadronic models,
dashed lines to hybrid models.}
\label{gwaves}
\end{figure}

A second indicator of EoS effects is the GW signal. Since the waveform is
sensitive to dynamical mass motions we expect that all the above
maximal density differences are reflected in the GW signal mainly in
the form of different frequencies. In Fig. \ref{gwaves} we plot the waveforms of all models sorted according to the initial mass.
Model E1/E2, the most massive one is the only one which shows a
significant pre-merger EoS difference in the waveform. This is because
the above mentioned maximal density difference translates via the
compactness and the binary angular velocity into a GW frequency
difference which amounts to $\sim10\%$ at the ISCO (cf. Fig. \ref{fig:ISCOfreq}) while it disappears during the actual merger phase.
The next less massive model, D1/D2, does not show any EoS differences
in the GW frequency at the ISCO and further during the pre-merger
phase, but we may expect, that for larger binary distances and
therefore smaller tidal interaction, GW frequency differences could be
seen. The more interesting part of this model is the different
collapse behaviour and therefore the totally different waveform. While
the hybrid model D2 only produces a short, high-frequent burst before
the collapse, the hadronic model emits a long wavetrain which
decreases slowly in amplitude. Models C1/C2 and B1/B2
emit both a quasi-periodic (QP) GW-signal which is characteristic for the
rotation and oscillation mode in the merger remnant. EoS-differences 
in the waveform do not appear until a considerable mixed matter core has
formed in the hybrid case. As a consequence, the first part of the
post-merger GW signal, which is the strongest in our simulations, will
not be affected by any EoS difference. However, when the mixed matter core becomes large enough, an
accumulating phase shift in the waveform becomes clearly visible. In
the C1/C2-model the shift adds up to more than half a period before
the collapse of the hybrid remnant happens, in the B1/B2 model, the shift is only very small, but visible.

\subsection{Gravitational Wave Spectra}

\begin{figure}
\includegraphics[height=21.5cm,width=8cm]{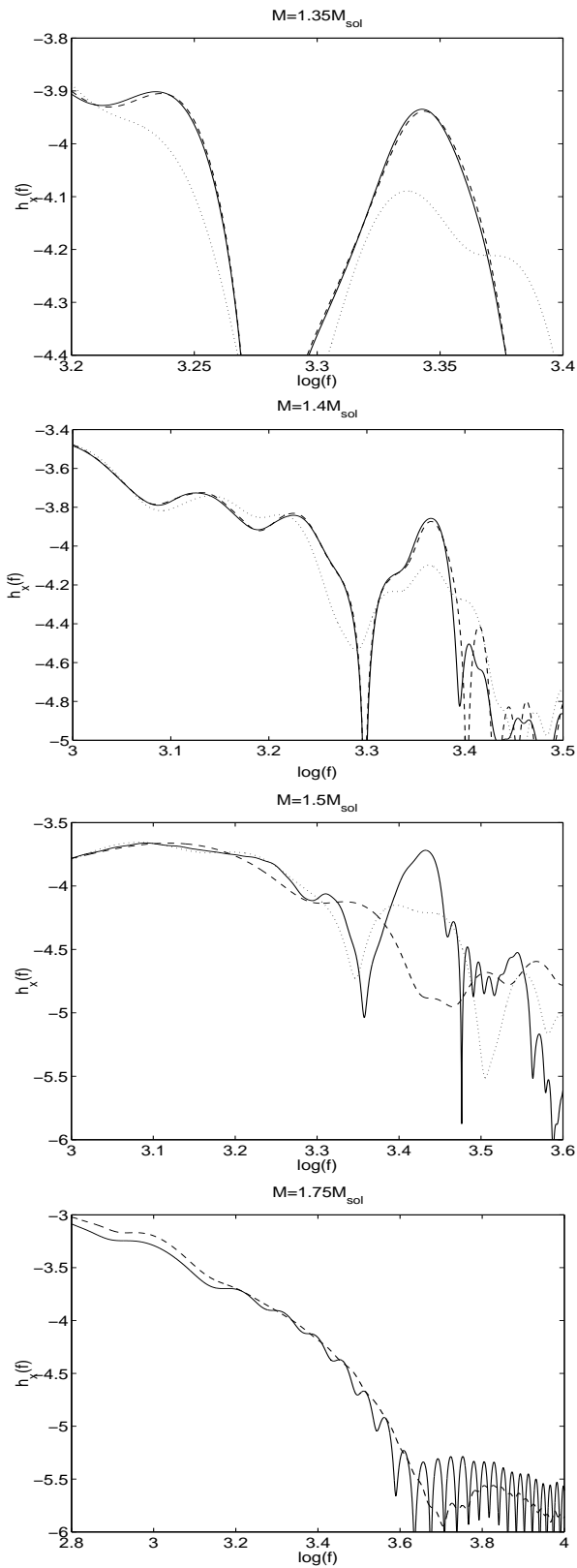}
\caption{Fourier spectrum of the waveform $h_+$ for models B1/B2
to E1/E2 sorted according to their initial mass. The frequency is in
units of kHz. Solid lines correspond to hadronic models, dotted lines
to the spectra of the corresponding truncated waveforms (see text) and
the dashed lines correspond to hybrid models.}
\label{fig:ftrafo}
\end{figure}

The Fourier spectrum of the GW waveform is plotted in Fig.\ref{fig:ftrafo}. 
The most massive E1/E2-model shows its EoS
differences in the pre-merger part, i.e in the frequency domain around
1kHz.  Here the spectrum produced by the pre-merger hybrid waveform is
significantly stronger than the one of the hadronic waveform, 
since the amplitude of wave strain $h$ scales roughly as 
$(r/M) h \sim (M/R)_\infty$ where $r$ is a distance from a source.  
This is consistent with simulations using quasi-equilibrium sequences
\citep{faberletter}. There is no contribution on the high frequency side from
this model since both remnants collapse immediately after
merger. Note that we miss the waveform emitted by the ringing of the 
resulting BH. However, the expected frequency of this signal lies in the
range of $5-10$kHz \citep{shibataPTP, leaver} we can clearly separate
out the BH waveform.

The other models show their EoS-differences via different frequency
peaks which result from the QP-waveform emitted in the
post-merger. The difference is most obvious in the D1/D2-model. For
the hadronic EoS model D1, a distinct Fourier peak is clearly visible,
whereas the hybrid EoS model D2 leads only to a much weaker and very
broad Fourier peak since the remnant is collapsing very soon after
merging producing a much shorter post-merger GW signal. In the C1/C2-model, both the hadronic and the hybrid models have strong QP
peaks where the one of the C2-model is slightly shifted to higher frequencies due to the more compact remnant. The
same can be seen for the model B1/B2. However, effects here
are smaller and a measurement may be difficult. The EoS-effects are
this tiny because the spectrum is dominated by the first GW burst
after merging which is insensible to EoS-differences.

We may compare the obtained QP frequencies $f_{QP}$ to values from a fully
relativistic calculation \citep{shibataPTP} where an irrotational
initial velocity field with a polytropic EoS is considered. Two peaks
are found in their spectrum
\begin{eqnarray}
f_{QP1}&\sim&1.8f_{QE}\\
f_{QP2}&\sim&2.8f_{QE}
\end{eqnarray}
for an adiabatic index of $\Gamma=2.25$ and
\begin{eqnarray}
f_{QP1}&\sim&1.8f_{QE}\\
f_{QP2}&\sim&3.0f_{QE}
\end{eqnarray}
for $\Gamma=2$.
Here, $f_{QE}$ is the GW frequency of the binary at the
ISCO. \citep{shibataPTP} point out, that $f_{QP2}/f_{QE}$ is not very dependent
on the initial compactness of the stars but the higher peak
$f_{QP2}/f_{QE}$ depends on $\Gamma$ in a way that softer EoSs lead to higher
values for $f_{QP2}/f_{QE}$. Since we use realistic EoSs with a
non-constant $\Gamma$ in our simulations, we expect the ratio
$f_{QP2}/f_{QE}$ to vary with increasing compactness \cite[][]{zang,faberII,oechslinPRD}. Indeed our values for
$f_{QP2}/f_{QE}$ are strongly dependent on the initial mass $M_\infty$
and compactness as can be seen in Tab. \ref{tab:qpfreq}. From the behaviour
of $f_{QP2}/f_{QE}$ we can deduce that the EoS softens for more
compact models, i.e. the EoS softens at higher densities. It might be
very interesting to calculate the $f_{QP2}/f_{QE}$-$(M/R)_\infty$-relation
for various other realistic EoSs on the market. Measuring this relation might
give a detailed insight to the NS EoS in the nuclear regime above $10^{14}$g/cm${}^3$.

One might expect that $f_{QP2}$ is overestimated in our simulations due
the already mentioned numerical viscosity effects in the code which
leads to a slowly contracting merger remnant and therefore to emission
of GW with higher frequencies. To check that this is not the case, we have also calculated the spectra of the truncated wavesignals which
only contain the first few oscillations of the post-merger signals. The
resulting QP peaks are reduced in strength but their positions remains
unchanged within the uncertainty of the peak width (see Fig. \ref{fig:ftrafo}).

\begin{table}
\caption{Ratio of $f_{QP2}$ to $f_{QE}$ for all models where a
transient NS forms or a delayed collapse happens. The values for
$f_{QE}$ are determined from quasi-equilibrium models and are taken at the
ISCO.}
\label{tab:qpfreq}
\begin{tabular}{lcccc}
\hline
Model & $M_\infty[M_\odot]$ & $C=(M/R)_\infty$ & $f_{QP2}$[kHz] & $f_{QP2}/f_{QE}$\\
\hline
A1&1.2&0.128&1.98&2.47\\
B1/B2&1.35&0.143&2.20&2.56\\
C1/C2&1.4&0.148&2.32&2.65\\
D1&1.5&0.158&2.70&2.97
\end{tabular}
\end{table}

\begin{figure}
\includegraphics[width=8.5cm]{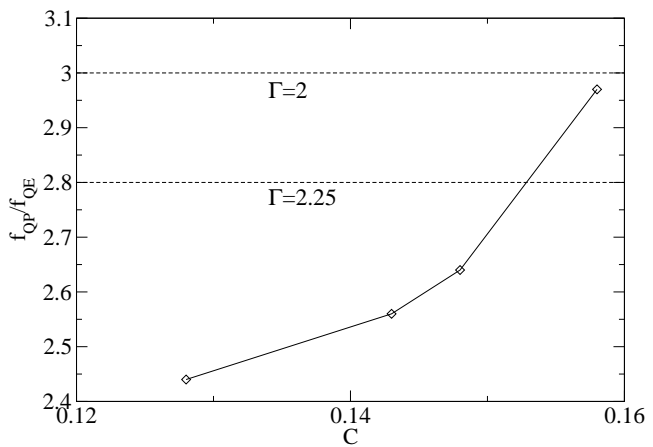}
\caption{Relation between the ratio $f_{QP2}/f_{QE}$ and the compactness
$C=(M/R)_\infty$. The diamonds correspond to the hadronic EoS-models
A1 to D1, the values for $\Gamma=2$ and $\Gamma=2.25$ are taken from \citep{shibataPTP}}
\label{fig:fqpfqe}
\end{figure}

\section{Conclusions}
\label{sect:conclusions}
In this work, we have considered the impact of the EoS to the inspiral and
merger dynamics of a binary neutron star coalescence. We have performed
 equilibrium sequence studies to investigate the GW frequencies around
the ISCO. Then, dynamical simulations of the
merging event using a 3-dimensional SPH code have been carried
out. We have chosen a hadronic EoS based on the relativistic mean field
approximation and a hybrid EoS which is obtained by combining the
hadronic EoS with a MIT-bag-model-EoS in the high-density regime above
$\rho_t=1.8\rho_0$.
The gravitational wave frequency at the ISCO depends strongly on the
initial mass of the binary. For a generic mass of $2\times1.4M_\odot$,
we find a GW frequency of around 900Hz for both EoSs. EoS
differences become important for models with initial gravitational masses 
larger than $M_\infty\sim1.5M_\odot$. At this mass range, the central 
density of the individual stars reach the phase transition $\rho_t$ 
included in the hybrid EoS model. At masses close to the
maximum gravitational mass of the hybrid EoS, $M=1.78M_\odot$, the relative
difference in $f_{GW}$ becomes as large as $\sim10\%$.\\

The maximum density evolution during the merger depends sensitively on
the fact whether the phase transition density is crossed and at which
time. For very massive models like models E1/E2 with an initial mass of
$M_\infty=1.75M_\odot$ this is already fulfilled by the individual
companion stars in the pre-merger phase. Less
massive models with masses below $M_\infty\lesssim1.5M_\odot$ cross this
threshold later in the post-merger phase when the matter is contracting to
form a transient NS or a BH. The hadronic EoS models all form
a transient NS except the very massive E1 model which collapses
immediately. In the D1 model, the maximum density does not converge
after 12ms but rises slowly and continuously. We think
that this is an effect of a considerable angular momentum
transport from the remnant core to the outer layers which
implies a continuous conversion of the differential rotation pattern
into uniform rotation.
All hybrid EoS models collapse on timescales
which is highly dependent on the model mass. While the massive models
E2 and D2 collapse immediately, models C2 and B2 collapse after a
contraction of the remnant on several dynamical timescales and a considerable decrease of the degree of
differential rotation. This indicates that the collapse is driven in
this case by angular momentum transport caused e.g. by numerical
viscosity. The collapse dynamics of these models will be investigated
in future work. 
The emitted gravitational waveforms are very dependent on the initial
mass of the model. The most massive model E1/E2 shows a large
difference prior to merger both in frequency and amplitude. This is a
consequence of the different compactnesses for the models E1 and E2 due to the
different EoSs. The same effect is not visible in the less massive
models because the compactness difference due to the EoSs vanishes.
A second aspect of the GW signal is the waveform emitted by the
remnant. If the remnant does not collapse immediately to a BH, a
quasi-periodic waveform with a frequency of 2-3kHz is emitted. The
ratio of this frequency to the GW frequency at the
ISCO $f_{QP2}/f_{QE}$ depends sensitively on the model and therefore on the
mass or compactness $(M/R)_\infty$. Since $f_{QP2}/f_{QE}$ is fairly
constant for an EoS with constant stiffness, we interpret this result
as a consequence of the varying stiffness of our EoS used. Hence,
the relation $f_{QP2}/f_{QE}$-$(M/R)_\infty$ is characteristic for the
behaviour of the EoS's stiffness. A measurement of this
quantity would provide important additional information of the EoS.

It has been suggested in \cite{Hughes} that measurements of high 
frequency GW will require a network of broadband detectors combined 
with narrowband detectors that have a good sensitivity in the 
high frequency domain. 
Such measurements of high frequency GW may become feasible in the 
future and discover or constrain the existence of phase transition 
in neutron star matter at high densities.

\vspace{1cm}
{\bf Acknowledgments}\\
\vspace{0cm}\\
Computations were done at the Department for Physics and Astronomy at
the University of Basel. RO would like to thank T. Janka for
discussions and helpful comments. KU would like to thank M. Shibata
and J. L. Friedman for discussion.\\
RO, GP and FKT are funded by the Schweizerischer Nationalfonds under grant 2000-061822.00. 

\bibliographystyle{mn2e.bst}

\end{document}